\documentclass[11pt]{article}
\usepackage{fullpage,amssymb,amsmath,graphicx,natbib,multirow,hyperref}

\renewcommand{\v}[1]{\boldsymbol #1}

\newcommand{\Exp}[1]{{\rm E}[ \ensuremath{ #1 } ]  }
\newcommand{\Var}[1]{{\rm Var}[ \ensuremath{ #1 } ]  }

\newcommand{\bbeta}{{\boldsymbol\beta}}
\newcommand{\blambda}{ {\boldsymbol\lambda}}

\title{Bayesian sandwich posteriors for pseudo-true parameters}
\author{ 
Peter Hoff and Jon Wakefield \\ 
       Departments of Statistics and Biostatistics\\
       University of Washington }
\date{\today}

\begin{document}
\maketitle

\let\thefootnote\relax\footnote{
This note is part of a discussion of ``Bayesian inference with misspecified models''
 by Stephen Walker.
Replication code for the simulation study is available at the first author's
website:
\href{http://www.stat.washington.edu/~hoff}{\nolinkurl{www.stat.washington.edu/~hoff}}. 
}

\begin{abstract}
Under model misspecification, the MLE generally 
converges to the pseudo-true parameter, the 
parameter corresponding to the 
distribution 
 within the model that is closest to the 
distribution from which the data are sampled.  
In many problems, the pseudo-true parameter 
corresponds to a population parameter of interest, and 
so a misspecified model can provide consistent 
estimation for this parameter. Furthermore, 
the well-known sandwich variance formula of \citet{huber_1967} 
provides an asymptotically 
accurate sampling distribution for the MLE, 
even under model misspecification. 
However, confidence intervals based on a 
sandwich variance estimate
may
behave poorly 
for low sample sizes, partly
due to the use of a plug-in estimate of the variance. 
From a Bayesian perspective, plug-in estimates of nuisance parameters
generally underrepresent 
uncertainty in the unknown parameters, and  
averaging over such parameters is  expected to 
give better performance. With this in mind, 
we present a Bayesian sandwich posterior 
distribution, whose likelihood 
is based on 
the sandwich sampling distribution of the MLE. 
This Bayesian approach allows for the incorporation of prior information 
about the parameter of interest, averages over uncertainty in the nuisance 
parameter and is asymptotically robust to model misspecification. 
In a small simulation study on estimating 
a regression parameter under heteroscedasticity, 
the addition of 
accurate prior information and the averaging over the nuisance parameter 
are both seen to improve the accuracy and calibration of 
confidence intervals for the parameter of interest. 

\medskip

\end{abstract}

\section{Introduction}

Let $X$ be the data resulting from an experiment, survey or 
random process, and let $\theta$ denote some fixed but unknown aspect of the 
the data generating process. Before the experiment is run, 
both $X$ and $\theta$ are uncertain. A subjective Bayesian uses 
probability to represent pre-experimental uncertainty in both 
$X$ and $\theta$, and 
Bayes' rule to represent uncertainty in $\theta$ after having observed 
$X$. 
One appealing aspect of the subjective Bayesian approach is that it is 
an internally consistent and rational way to update information. 
If  $\mathcal P_{\Theta} = \{ p(X|\theta):\theta\in \Theta\}$ 
expresses 
our beliefs about $X$ given $\theta$ , and 
$\pi(\theta)$ expresses our beliefs about $\theta$, 
then $\pi(\theta|X) \propto \pi(\theta) p(X|\theta)$ expresses 
what we \emph{should} believe about $\theta$, having observed $X$. 
For $\pi(\theta|X)$ to be of 
most use, both 
$p(X|\theta)$ and $\pi(\theta)$ should actually represent our 
beliefs, at least approximately. 
Professor Walker's paper \citep{walker_2013} highlights the problem that in practice, 
a statistical model 
$\mathcal P_{\Theta}$ 
is often used that 
is known  to not represent beliefs, in that it is suspected that 
$\mathcal P_{\Theta}$ does  not include the distribution that 
generated the data. 
In such cases, interpretation of $\pi(\theta|X)$ may be problematic: 
Not only is the validity of $\pi(\theta|X)$ as 
a probabilistic description of information about $\theta$ potentially 
invalid, it is not even clear that $\theta$ represents anything of 
interest. 

One remedy discussed by Walker is to expand the model so 
that $\mathcal P_{\Theta}$ can be assumed to contain the correct 
data generating process, or at least something very close to it. 
Depending on what the data are, this can make the model 
quite large. Walker focuses on the situation where the data are 
taken to be a sample of observations from a population, 
i.e.\ $X= \{ x_1,\ldots, x_n\}$. To guarantee that the model is not 
misspecified, $\mathcal P_{\Theta}$ must be quite large, essentially
covering (in a topological sense) the space of all probability 
distributions. However, addressing the model misspecification problem in this 
way can complicate the  
other component of subjective Bayesian  inference - specification of the prior distribution. The larger the model is, the more difficult it will be to 
specify a prior  
that represents actual beliefs about the unknown population. 
If $\pi(\theta)$ does not represent prior beliefs, then 
the use of  $\pi(\theta|X)$  as an expression of 
posterior beliefs is questionable, except possibly when the sample size is 
very large. 

\section{Incorrect models with correct pseudo-true parameters}

If we wish to benefit from the internal consistency of 
subjective Bayesian inference, we need to limit our probability statements 
to those quantities about which we have actual information. 
As a very simple example, 
suppose we have a sample $x_1,\ldots, x_n$ of 
independent measurements for which the measurement error $\sigma^2$ 
is known. If we have prior information $\pi(\theta)$ about the 
population mean $\theta$, but not any other aspect of the 
population (other than $\sigma^2)$, 
then we should limit our 
data $X$ to quantities whose sampling distribution depends only 
on $\theta$ and  $\sigma^2$. This condition will 
be approximately  met  by the sample mean $\bar x$, 
whose sampling distribution 
is approximately normal. 
A limited form 
of subjective Bayesian inference can proceed via 
the posterior density  $\pi( \theta |\bar x)   \propto 
   \pi(\theta) \times p(\bar x|\theta)$, 
where the latter density is that of a $N(\theta,\sigma^2/n)$ random
variable. 

Strictly speaking, the model $p(\bar x|\theta)$ is misspecified 
unless $x_1,\ldots, x_n$ are sampled from a normal population. 
As \citet{walker_2013} asks, what does $\theta$ represent in the case of model misspecification?  Letting $p_0(\bar x)$ be the true sampling distribution of $\bar x$, 
the pseudo-true parameter $\theta^*$ is given by 
\begin{align*}
\theta^* &= \arg\min_{\theta}\int \log \frac{ p_0(\bar x) }{p(\bar x|\theta) } 
     p_0(\bar x) \ d\bar x   \\ 
&=  \arg\min_{\theta}  \int \tfrac{1}{2} [ 
   \log(2\pi \sigma^2/n) + n\bar x^2/\sigma^2 - 2n\bar x\theta/\sigma^2 + 
    n\theta^2/\sigma^2 ] p_0(\bar x) \ d\bar x  \\
&=  \arg\min_{\theta}   ( \theta^2 - 2 \theta \Exp{\bar x} ) 
= \Exp{\bar x } = \theta, 
\end{align*}
and so in this case, 
the pseudo-true parameter is equal to the parameter of interest, 
regardless of whether or not $p_0$ is in the model. 
Furthermore, the posterior distribution given by 
$\pi( \theta |\bar x)   \propto 
   \pi(\theta) \times p(\bar x|\theta)$ 
provides (approximate) subjective Bayesian inference for the population 
mean $\theta$, even if the population is not normal, and without 
having to quantify prior information about anything but the first two 
population moments. 

Now suppose we are interested in estimating a collection of 
population moments $\blambda\in \mathbb R^p$, 
where $\lambda_j =  \Exp{ g_j(x) }$,  $j=1,\ldots, p$. 
Is there a parametric model
$\{ p(x|\theta) : \theta\in \Theta\}$ 
 whose pseudo-true parameter $\theta^*$ satisfies 
$\Exp{ g_j(x)|\theta^*} = \lambda_j$ for each $j=1,\ldots, p$?
Consider the exponential family with sufficient 
statistics $\{ g_1(x) ,\ldots, g_p(x) \}$ given by 
$ p(x|\theta )
 = h(x) \exp\{ \theta_1 g_1(x) + \cdots \theta_p g_p(x) - c(\theta)\}$. 
The pseudo-true parameter $\theta^*$ for such a model is given by 
\begin{eqnarray*}
\theta^*  &=& \arg \min_\theta \int \log \frac{p_0(x)}{p(x|\theta)} p_0(x) 
  \ dx \\
&=& 
 \arg \max_\theta \int [ \log p(x|\theta) ] p_0(x) 
  \ dx \\
&=&\arg \max_\theta 
   \int [ \theta_1 g_1(x) + \cdots \theta_p g_p(x) ] p_0(x) \ dx- 
   c(\theta), 
\end{eqnarray*}
where $p_0(x)$ is the true population density. 
Taking derivatives with respect to each element of $\theta$ tells us that
$\theta^*$ is the solution in $\theta$ to
  \[ \frac{ \partial}{\partial \theta_j  }
  \int [ \theta_1 g_1(x) + \cdots \theta_p g_p(x) ] p_0(x) \ dx
    =   \frac{ \partial}{\partial \theta_j  } c(\theta).  \]
The left-hand side is $\int g_j(x) p_0(x) \ dx = \lambda_{j}$, one of the 
moments we want to estimate. The right-hand side is
equal to $\Exp{g_j(x)|\theta}$, due to the well-known identity for exponential 
families. 
Therefore, $\theta^*$ is the parameter value such that
$\int g_j(x) p(x|\theta^*) \ dx = \int g_j(x) p_0(x) \ dx$
for $j\in \{1,\ldots, p\}$.
Thus for an exponential family
with sufficient statistic $\{g_1(x),\ldots, g_p(x) \}$, 
the pseudo-true parameter  $\theta^*$ 
satisfies $\Exp{ g_j(x) |\theta^* } = \Exp{ g_j(x) }$, 
where the latter expectation is with respect to the true population distribution. 

The result above suggests that some models can 
be used to make inference for certain aspects of a population $P_0$, 
even if $P_0$ is not a member of the model. 
Specifically, a possibly incorrect model $\{ P_\theta : \theta\in \Theta\}$
can be used to obtain consistent estimators of those functionals of $P_0$ 
which match those of $P_{\theta^*}$, where $\theta^*$ is 
the pseudo-true parameter.  
However, 
this does not ensure that the model can correctly represent 
the sampling variability of such estimators, even asymptotically. 
As a result, 
confidence intervals based on an incorrect model 
can be asymptotically invalid, even if the incorrect model 
provides a consistent estimator. 
To address this concern, 
\citet{huber_1967} derived the limiting distribution of the MLE $\hat \theta$ 
of 
$\theta$ under a possibly incorrect model in terms of the pseudo-true parameter. The approximation 
proceeds roughly as follows: 
Suppose $x_1,\ldots, x_n$ are i.i.d.\  observations from population $P_0$, and 
let $l(\theta:x_i) = \log p(x_i|\theta)$ be the log-likelihood
corresponding to a single observation $x_i$. 
A first order Taylor series expansion of 
$ \sum  \dot l(\theta^*:x_i) $ around the MLE $\hat \theta$ gives
\[ \sum_{i=1}^n \dot l(\theta^*:x_i)  \approx \left ( \sum_{i=1}^n 
 \ddot l(\hat \theta :x_i ) \right ) (\theta^* -\hat \theta).  \]
By the central limit theorem, the sum on the left-hand side is 
approximately $N(0 ,n  B )$, where  $B= 
  \Var{ \dot l(\theta^*:x) }$ and the variance here is under $P_0$. 
Letting $A= \sum_{i=1}^n \ddot{l}(\hat \theta :x_i) $ be the sum on the 
right-hand side, we have 
\begin{equation}
   (\theta^* - \hat \theta) \   \dot\sim  \
     N( 0 ,n  A^{-1} B A^{-1}  ), 
\label{eq:sna}
\end{equation}
where ``$\dot\sim$'' means  ``approximately distributed as.''
This result has been used extensively to obtain
confidence intervals for the pseudo-true parameter $\theta^*$, in cases 
where it corresponds to a population quantity of interest. 
In practice, since $\theta^*$ is unknown, $B$ is 
estimated as $\hat B = \sum  \dot l(\hat \theta:x_i)  \dot l(\hat \theta:x_i)^T/n$, 
the sample variance of the likelihood functions at the MLE. 
The resulting variance estimate $\hat C = n A^{-1} \hat B A^{-1}$ 
is called the sandwich variance estimate for $(\theta^*-\hat \theta)$. 
Confidence intervals for $\theta^*$ can be obtained by 
approximating the distribution of $\hat C^{-1/2} (\theta^* - \hat \theta)$
by a  $N(0,I)$ distribution. 
Sandwich confidence intervals avoid the issue of model misspecification 
by positing the sampling distribution of the pivotal quantity
 $\hat C^{-1/2} (\theta^* - \hat \theta)$, rather than the sampling 
distribution of $x_1,\ldots, x_n$. The model used to obtain the 
likelihoods $\{ l(\theta:x_i),i=1,\ldots, n\}$ is simply a tool 
that provides a consistent estimate of the pseudo-true parameter 
$\theta^*$ and asymptotically correct confidence intervals via 
the sandwich variance estimate. 
A review of the theory and methods for 
sandwich-based data analysis appears in \citet{white_1982}, and 
several applications are described in \citet{white_1980} and 
\citet{royall_1986}. 
Sandwich variance estimation has also been applied  to 
inference based on generalized  estimating equations (GEE), 
a popular likelihood-free approach to  inference
\citep{liang_zeger_1986,zeger_liang_1986,gourieroux_montfort_trognon_1984}.


\section{A Bayesian sandwich posterior distribution}

While used extensively in practice, sandwich confidence intervals 
can behave poorly for low sample sizes, 
with coverage often being 
well below their  nominal level \citep{kauermann_carroll_2001}. 
One reason for this is that 
the sandwich procedure does not properly account for 
uncertainty in the variance
  $B$ of $\dot l(\theta^*:x)$. 
The replacement of $B$ by 
 $\hat B$ in fact uses two plug-in approximations: the MLE $\hat \theta$ 
for $\theta^*$, and the sample covariance $\hat B$ for the population 
covariance $B$. Ignoring the uncertainty in both of these approximations is 
likely to provide an underestimate 
of $B$, resulting in overly-narrow confidence intervals 
and below-nominal coverage rates. 

One of the attractions of Bayesian inference is that uncertainty 
in nuisance parameters can be accounted for by integrating 
over their possible values, rather than plugging in point estimates. 
With this in mind, we propose the following version of 
a ``Bayesian sandwich'' posterior distribution, 
quantifying the uncertainty in both $\theta^*$ and $B$:
Given a 
working model $\mathcal P_{\Theta} =\{ p(x|\theta) : \theta\in \Theta\}$
and 
observations $x_1,\ldots, x_n \sim $ i.i.d.\ $P_0$, 
we form a likelihood derived from the approximate joint density of the 
MLE $\hat \theta$ based on 
$\mathcal P_{\Theta}$, and the sum of squares of the  derivatives of the 
 log-likelihood functions
$S(\theta) = \sum_{i=1}^n  \dot l(\theta:x_i)  \dot l(\theta:x_i)^T$,
giving us the following approximate likelihood function:
\begin{align*}
  p(\hat \theta , S(\theta^*) | \theta^*,B)   &=
    p(\hat \theta | \theta^*,B) \times 
   p(S(\theta^*) | \hat \theta, \theta^*,B) \\
 &\approx {\rm dnorm}(\hat \theta   | \theta^*, n A^{-1} B A^{-1}) \times 
  {\rm dWishart}(S(\theta^*)   | n, B), 
\end{align*}
where ``dnorm'' and ``dWishart'' refer to the normal
and Wishart densities respectively. 
The validity of this likelihood is based on three approximations. 
The first is the  normal approximation to the distribution of $\hat \theta$ given by 
$(\ref{eq:sna})$.  The second is the conditional independence of 
$S(\theta^*)$ and $\hat \theta$ and the third is the approximation 
of the distribution of $S(\theta^*)$ with a Wishart distribution. 
The first of these approximations is justified asymptotically 
by \citet{huber_1967}, 
whereas the latter two are, at least currently, heuristic. 

Based on this approximate likelihood
and a prior distribution 
for $(\theta^*,B)$, a posterior distribution can be obtained 
via MCMC in the usual way. For example, if the priors for 
$\theta^*$ and $B$ are normal$(m_0,V_0)$ 
and inverse-Wishart$(\nu_0,S_0^{-1})$ respectively, 
then posterior approximation can proceed via the following 
Gibbs sampler: Given current values of  $\theta^*_{(s)}$ and $B_{(s)}$, 
\begin{enumerate}
\item simulate $\theta^*_{(s+1)} \sim N_p ( m_1 ,V_1)$, where 
\[ 
V_1^{-1} =  V_0^{-1} + A B_{(s)}^{-1} A /n  \ \ , \ \ 
m_1 = V_1[  V_0^{-1} \theta_0  + A  B_{(s)}^{-1} A \hat \theta/n   ], \]
\item simulate $B_{(s+1)}^{-1}  \sim $ Wishart$(\nu_1 ,S_1^{-1})$ , where  
$\nu_1=\nu_0 +n +1$ and 
\[  S_1 =  S_0 + S(\theta^*_{(s+1)} ) + A (\theta^*_{(s+1)} - \hat \theta  ) 
         (\theta^*_{(s+1)} - \hat \theta  )^T A /n.   \]
\end{enumerate}

The hyperparameters $(m_0, V_0)$ should ideally represent prior 
information 
about $\theta^*$. Information upon which to base $(\nu_0,S_0^{-1})$
might be harder to come by. 
One 
possibility would be to use Jeffreys' prior, $\pi(B) \propto |B|^{-(p+1)/2}$  
\citep{geisser_cornfield_1963}. The posterior distribution under 
this prior can be approximated with the above Gibbs sampler by setting 
$\nu_0=0$ and $S_0$ equal to the $p\times p$ matrix of zeros. 


\section{Example: Regression with heteroscedastic errors}
Suppose we have a sample  $(y_1, x_1),\ldots, (y_n,x_n) \sim$ i.i.d.\ 
 $P_0$, and wish to estimate the linear regression of $y$ on $x$ 
that would be obtained from performing the regression 
 on the entire population. In other words, letting $\v x =(1,x)$, we 
want to estimate 
$\bbeta = \Exp{ \v x \v x^T }^{-1} \Exp{ \v x y }$, 
where both expectations are under $P_0$. 
Consistent estimation of this quantity can be obtained from 
the  normal regression model 
$y_i = \bbeta^T\v x_i + \epsilon_i$,  $\epsilon_1,\ldots, \epsilon_n
\sim $ i.i.d.\ $N(0,1)$, even if $P_0$ is not in this model, 
 as  the pseudo-true parameter of this regression model 
is equal to  $\Exp{ \v x \v x^T }^{-1} \Exp{ \v x y }$, 
the population regression parameter under $P_0$. 
We also note that the variance of the error terms in the 
regression model can be taken to be any fixed value: Whichever value 
is specified will end up canceling out in the 
sandwich variance calculation. 

Using the regression model as our working model, we have 
 $\dot l(\bbeta:y,\v x) = \v x(y- \bbeta^T\v x)$ and 
 $\ddot l(\bbeta:y,\v x) = -\v x\v x^T$, giving 
\[ 
A =  - \sum_{i=1}^n \v x_i \v x_i^T \  \ \mbox{and}  \ \ 
 S(\bbeta) = \sum_{i=1}^n \v x_i(y_i- \bbeta^T\v x_i).
\]
The usual sandwich 
variance estimate of the  MLE $\hat{\bbeta}$ under the 
normal regression model is $n A^{-1} \hat B A^{-1}$
where $\hat B = S(\hat \bbeta)/n$. In contrast, 
the proposed Bayesian sandwich 
posterior distribution infers $B$ 
jointly with 
$\bbeta$, based on the Wishart model for $S(\bbeta)$. 
To compare the performance of the proposed  Bayesian sandwich posterior 
to the usual sandwich procedure, we ran a small simulation study 
in order to calculate  coverage
rates and average interval widths 
 of nominal 95\% confidence 
intervals.
For both  
 small ($n=10$) and  large ($n=500$) sample sizes, 
datasets were generated as 
$x_1,\ldots, x_n \sim $ i.i.d.\ exponential(1), and 
$y_i|x_i \sim N( \beta_{1} + \beta_{2} x_i, (\beta_{1}+\beta_{2} x_i)^2 )$, 
where $\beta_{1}=\beta_{2}=1$. 
Thus the working model incorrectly assumes homoscedastic errors, 
whereas the true population has substantial heteroscedasticity. 

For each simulated dataset, we obtained Bayesian sandwich posterior 
distributions under 
four different priors of the form $\pi(\bbeta,B) = \pi(\bbeta) \pi(B)$, 
based on two choices for each of $\pi(\bbeta)$ and $\pi(B)$. 
The priors for $\bbeta$ included 
the (improper) uniform prior on $\mathbb R^2$, and an 
informative $ N( (1,1)^T , nA^{-1} )$ prior distribution. 
This latter prior, weakly centered around the correct values, 
represents accurate but weak information
about $\bbeta$ that someone may have: The matrix $A =\sum \v x_i \v x_i^T$  is the information for $\bbeta$ from $n$ observations, and so
$A/n$ represents  the information equivalent of one observation.
The priors for $B$ included  Jeffreys' prior and a point-mass prior on 
$\hat B$, the plug-in estimate of $B$. 
We note that the uniform/plug-in combination of priors 
leads to a $N_p(\hat \bbeta, nA^{-1}\hat B A^{-1})$ posterior 
distribution for $\bbeta$. This posterior was referred to as 
the ``artificial posterior'' by \citet{muller_2011}, who compared 
the risk of the resulting estimator  to the risk of the Bayes 
estimator  from the working model.

For each sample size we simulated 10,000 datasets from the heteroscedastic 
regression distribution given above, and obtained  95\%  posterior
confidence  intervals for the slope  $\beta_2$ based on each of the 
four priors.
Empirical coverage probabilities and average
interval widths are given in Table \ref{tab:res}.
For each dataset 
we also obtained a Wald-type interval 
for $\beta_2$ based on the plug-in sandwich variance estimate 
(the usual sandwich confidence interval), 
but it performed nearly identically to the 
estimator based on the uniform/plug-in prior, so we do not report these results separately. 
\begin{table}
\begin{center}

\begin{tabular}{ c c | c c }  
 &     & \multicolumn{2}{c}{ $\pi(\bbeta)$ } \\  
 &  $n=10$                          & informative & uniform \\ \hline
\multirow{2}{*}{$\pi(B)$} &Jeffreys & 0.95 (2.86)   & 0.87 (4.80)  \\
                         & plug-in  & 0.69  (2.14)  & 0.65 (2.63)  
\end{tabular}  \ \ \ 
\begin{tabular}{ c c | c c }
 &     & \multicolumn{2}{c}{ $\pi(\bbeta)$ } \\
 &  $n=500$                         & informative & uniform \\ \hline
\multirow{2}{*}{$\pi(B)$} &Jeffreys &  0.94 (0.74) & 0.94 (0.76) \\
                         & plug-in  &  0.93 (0.73) & 0.93 (0.74)  
\end{tabular} 
\end{center}
\caption{Coverage rates and average interval widths (in parentheses)  of 10,000
 nominal 95\% confidence intervals based on the four procedures.  
 The standard (non-Bayesian) sandwich procedure corresponds closely 
to the uniform/plug-in prior combination.}
\label{tab:res}
\end{table}

For $n=10$, both plug-in procedures perform very poorly 
in terms of coverage. This seems primarily due to underestimation
of $B$, resulting in confidence intervals that are 
shorter than are required to attain 95\% coverage. 
In contrast, the  procedures using Jeffreys' prior
both take uncertainty in $B$ into account, and provide 
coverage rates closer to the nominal value. However, the absence of 
any prior information about $\bbeta$  (uniform $\pi(\v \beta)$) 
leads to 
interval widths that are quite high as compared 
to those obtained with some prior information (informative $\pi(\v \beta)$),
as we would expect:
Accurate  prior information about $\bbeta$ leads to more precise inference. 
For $n=500$, 
all sandwich-based procedures performed similarly, 
reflecting the asymptotic correctness of sandwich-based
confidence intervals in general.
This is in contrast to the 
95\% nominal 
posterior confidence intervals based on the (uncorrected) misspecified 
regression model. For a sample size of $n=500$ and under the 
informative prior described above, 
these  95\% posterior confidence intervals had 
a coverage rate of only  68\% .

\section{Discussion}
Bayesian inference typically proceeds via the formulation of a 
sampling model 
for the data $X$ and a prior distribution 
over the sampling 
model. To guard against model misspecification, one approach 
is to make the model large enough to ensure that it contains 
the distribution that generated the data. However, such a large model can
lead
to difficulties in prior specification and posterior calculation. 
Such difficulties can often be avoided  when interest is 
limited to 
a simple low-dimensional parameter $\theta$.  
In such cases there often exists a 
statistic $t(X)$ 
or  pivotal quantity $s(X,\theta)$ whose sampling distributions 
are robust to model misspecification and 
from which a likelihood can be constructed. 
In this note, we have suggested using the asymptotic ``sandwich'' distribution 
of the MLE to construct a likelihood, and have 
illustrated via simulation how
Bayesian confidence intervals based on this likelihood  
provide improved performance over the standard non-Bayesian 
procedure. Other authors have used similar ideas previously: 
In a testing context, 
\citet{johnson_2005} shows how modeling the distribution of test statistics, 
rather than the individual observations, can 
lead to great simplifications in the calculation of Bayes factors 
(see also \citet{wakefield_2009}). 
In a semiparametric 
estimation setting, \citet{hoff_2007a} proposes Bayesian inference 
via a  marginal likelihood that depends only on the parameter of 
interest and not an infinite-dimensional nuisance parameter. 
Approaches  such as these suggest that simple,  robust
Bayesian inference can be obtained by 
restricting attention to 
only those aspects of the data for which
confident probability statements can be made.


\bibliographystyle{chicago}
\bibliography{/Users/hoff/Dropbox/SharedFiles/refs}

\end{document}